# Wind-Swept Clouds and Possible Triggered Star-Formation associated with the Supernova Remnant G357.7+0.3

J.P. Phillips, G. Ramos-Larios & J.A. Pérez-Grana

Instituto de Astronomía y Meteorología, Av. Vallarta No. 2602, Col. Arcos Vallarta, C.P. 44130 Guadalajara, Jalisco, México   e-mail : jpp@astro.iam.udg.mx

**Abstract**

We present evidence for interaction between the supernova remnant (SNR) G357.7+0.3 and nearby molecular clouds, leading to the formation of wind-swept structures and bright emission rims. These features are not observed at visual wavelengths, but are clearly visible in mid-infrared (MIR) mapping undertaken using the Spitzer Space Telescope (SST). Analysis of one of these clouds, the bright cometary structure G357.46+0.60, suggests that it contains strong polycyclic aromatic hydrocarbon (PAH) emission features in the 5.8 and 8.0 $\mu$m photometric bands, and that these are highly variable over relatively small spatial scales. The source is also associated with strong variations in electron density; an FIR continuum peak associated with dust temperatures of ~ 30 K; and has previously been observed in the 1720 MHz maser transition of OH, known to be associated with SNR shock excitation of interstellar clouds. This source also appears to contain a YSO within the bright rim structure, with a steeply rising spectrum between 1.25 and 24 $\mu$m. If the formation of this star has been triggered recently by the SNR, then YSO modelling suggests a stellar mass ~ 5-10 $M_\odot$, and luminosity $L_{YSO}$ ~ $10^2 \rightarrow 210^3$ $L_\odot$.

Finally, it is noted that a further, conical emission region appears to be associated with the Mira V1139 Sco, and it is suggested that this may represent the case of a Mira outflow interacting with a SNR. If this is the case, however, then the distance to the SNR must be ~ half of that determined from CS J=2-1 and 3-2 line radial velocities.

**Key Words:** stars: formation --- (stars:) supernovae: individual: G357.7+0.3 --- ISM: kinematics and dynamics --- ISM: jets and outflows --- ISM: clouds --- (ISM:) supernova remnants

# 1. Introduction

The formation of stellar clusters is known to radically affect the structure and energetics of nearby interstellar material, and the clouds out of which the clusters were born. Massive stars possess winds with velocities in excess of ~$10^3$ km s$^{-1}$, for instance, corresponding to energy losses of > $10^{36}$ erg/s (e.g. Preibisch & Zinnecker 2001), and are capable of disrupting parental clouds at a rate of ~$10^{-2}$ M$_\odot$ yr$^{-1}$ (cf. Yorke 1986). A $10^4$ M$_\odot$ molecular cloud can be disrupted within ~ $10^6$ years. Similarly, local molecular clouds are likely to be composed of condensations which are relatively small, with masses ranging down to ~$10^{-4}$ M$_\odot$, and these will be photoionised by B0-O4 stars out to a distance of ~30 pc from the cluster providing that the mean density in the region is ~10 cm$^{-3}$ (McKee, Van Buren & Lazareff 1984). One sided photoionisation of the larger clumps is also capable of radially accelerating them to velocities > 5 km s$^{-1}$ (the "rocket" effect), and this is sufficient to clear regions of size greater than ~ 15 pc (Bertoldi & McKee 1990).

Finally, stars with mass > 8 M$_\odot$ will be expected to transform into supernovae over periods of between 3 and 30 Myrs, and even relatively sparse clusters with 5 or so massive stars can be expected to witness a supernova explosion within 3 Myrs years of their formation (Shull & Saken 1995; Hester & Desch 2005). The massive Scorpius-Centaurus OB association, which has formed within the last 8-12 Myr (Preibisch & Zinnecker 1999; Mamajek, Meyer, & Liebert 2002), may have witnessed ~20 supernovae at intervals of ~5 $10^5$ years (Maíz-Appellániz 2001).

The shock waves of such supernovae have initial velocities of > $10^4$ km s$^{-1}$, and transfer ~$10^{51}$ ergs of energy to the ambient ISM (Preibisch & Zinnecker 2001). Studies of the interaction of these shock waves with molecular clouds (e.g. Boss 1995; Foster & Boss 1996, 1997; Vanhala & Cameron 1998; Fukuda & Hanawa 2000) show that clouds are shredded and destroyed where velocities are greater than ~ 50 km s$^{-1}$ - a situation which is likely to occur within ~10 and 100 pc of the supernova event, and depends upon the structure of the surrounding medium, as well as the evolutionary state of the pre-impact cores (e.g. Oey & García-Segura 2004; Vanhala & Cameron 1998). The lower velocities which prevail outside of this regime lead to shocking of the neutral condensations, local gravitational instability, and the development of further phases of so-called "triggered" star formation (e.g. Boss 1995; Foster & Boss 1996, 1997; Nakamura et al. 2006; Melioli et al. 2006; Boss et al. 2008).

There is now a significant amount of evidence for shock interaction between SNR and molecular shells. The shells are usually characterized by broad CO, OH and other molecular transitions (see e.g. Arikawa et al. 1999, Reach et al. 2005 (W28); Reach & Rho 1999, Reach et al. 2002 (3C 391); Seta et al. 1998, Reach, Rho & Jarrett 2005 (W44); Koo et al. 2001, Byun et al. 2006 (HB 21); Koo & Moon 1997 (W51C), and the well-known case of IC 443 (Burton 1987; Dickman et al. 1992; Tauber et al. 1994; Turner et al. 1992; van Dishoeck, Jansen & Phillips 1993; Snell et al. 2005)), lines which are presumably created through re-combination behind dissociative shocks, or as a result of slow acceleration in non-dissociative shocks (Koo 2003). Such gas also appears to be warmer and denser than is observed in most molecular clouds, and this leads to ratios between high and lower frequency CO rotational transitions which are greater than unity.

Such regions are also associated with the shock excitation of $H_2$ and other molecules (see e.g. Burton et al. 1988; Richter, Graham & Wright 1995; Rosado et al. 2007), and enhanced emission from the 1720 MHz ($^2\Pi_{3/2}$, J = 3/2, F = 2→1) maser line of the OH molecule (e.g. Wardle & Yusef-Zadeh 2002, and references therein). These latter transitions arise in slow non-dissociative C-shocks with column densities $10^{16}$-$10^{17}$ cm$^{-3}$; a circumstance which is likely to arise where the shocks are observed tangentially to the line of sight (Elitzur 1976; Lockett et al. 1999; Koo 2003).

We report the detection of a further region of likely shock interaction between interstellar clouds and a supernova remnant. In this present case, a region of radially wind-blown features, observed using the Spitzer Space Telescope (SST), is shown to be located at the limits of the SNR G357.7+0.3 discovered by Reich & Furst (1984). These globules possess bright-rimmed structures on the sides facing the remnant, presumably arising from shock interaction with the outflowing SNR gas, and tails which appear to be pointing directly away from the centre of the SNR. Our results also reveal evidence for Class I young stellar objects (YSOs) throughout the region, associated (in most cases) with the wind-blown shells. These YSOs are likely to have been triggered by the SNR.

We shall discuss the broad morphology of this region, and the detailed properties of one of the more interesting of the cometary structures. We shall also discuss the properties of a likely YSO located within the bright rim of this globule.

Finally, we shall take note of a conical emission region associated with the Mira V1139 Sco. It is suggested that we may be observing an interaction between the Mira and the SNR, leading to the acceleration of the Mira wind to large distances (> 14 pc) from the star.

## 2. Observations

We shall be making use, in the following analysis, of data products deriving from the near infrared (NIR) 2MASS all sky survey; the far infrared (FIR) survey undertaken with the Infrared Astronomical Satellite (IRAS); the second Galactic Legacy Infrared Midplane Survey Extraordinaire (GLIMPSE II), and the MIPSGAL project, both of the latter undertaken using the SST; and MIR photometry and imaging undertaken during the Mid-Course Space Experiment (MSX). Details of the data bases employed, and procedures used in the analysis of the data can be found in Skrutskie et al. (2006) (2MASS); at the web site http://irsa.ipac.caltech.edu/IRASdocs/toc.html (IRAS); in Benjamin et al. (2003) (GLIMPSE); in Carey et al. (2009) (MIPSGAL); and in Mill et al. (1994) and Price et al. (1998) (MSX). We shall describe here only the most salient features of importance to this analysis.

The catalogues and mosaics of the GLIMPSE II project were published over a period extending between 2007 and the Spring of 2008, and cover approximately 25 square degrees of the Galactic plane within the regimes $350° < \ell < 10°$, and $-2° < b < +2°$. Mapping was undertaken using the Infrared Array Camera (IRAC; Fazio et al. 2004), and employed filters having isophotal wavelengths (and bandwidths $\Delta\lambda$) of 3.550 μm ($\Delta\lambda$ = 0.75 μm), 4.493 μm ($\Delta\lambda$ = 1.9015 μm), 5.731 μm ($\Delta\lambda$ = 1.425 μm) and 7.872 μm ($\Delta\lambda$ = 2.905 μm). The spatial resolution varied between ~1.7 and ~2 arcsec (Fazio et al. 2004), and is reasonably similar in all of the bands, although there is a stronger diffraction halo at 8 μm than in the other IRAC bands. This leads to differences between the point source functions (PSFs) at ~0.1 peak flux. The maps were published at a resolution of 0.6 arcsec/pixel.

We have used 8 μm mosaic images from this survey to investigate broad regions of sky close to SNR G357.7+0.3 We have also produced contour mapping of the cometary structure G357.46+0.60 in all four of the IRAC bands, where contour levels are set at logarithmic intervals, and take the values $I_n = A10^{(n-1)B}$. The lowest level $n = 1$ corresponds to the outermost contour, whilst details of the parameters (*A, B*) are provided in the captions to the figures. We have also used the point source catalog (PSC) to identify likely young stellar objects (YSOs) within the areas surveyed in this present analysis.

The 3.6 μm fluxes for a YSO within the rim feature of G357.46+0.60 were acquired through small aperture photometry of the FITS image of the source. Levels of background were estimated through measures at a variety of locations about the source, and these were then subtracted from the original photometry to yield a measure of the uncontaminated YSO flux.

We have also obtained contour mapping of 8.0μm/4.5μm and 5.8μm/4.5μm flux ratios for the head-region of the cloud. This was undertaken by estimating levels of background emission, removing these from the 4.5, 5.8, and 8.0 μm images, and subsequently setting values at $< 3\sigma_{rms}$ noise levels to zero. The maps were then ratioed on a pixel-by-pixel basis, and the results contoured using standard IRAF programs.

Finally we have also determined overall fluxes for the cometary structure G357.46+0.60. This involved large aperture measures of the source, and equivalent measures at a variety of locations about the cloud. Both in this case, and for the flux ratio mapping, some care must be taken in evaluating appropriate levels of flux, however. The problems with large aperture photometry are described in the IRAC data handbook (http://ssc.spitzer.caltech.edu/irac/dh/iracdatahandbook3.0.pdf), and relate in part to scattering in an epoxy layer between the detector and multiplexer (Cohen et al. 2007). This leads to the need to modify fluxes as described in Table 5.7 of the handbook; corrections which are of maximum order 0.944 at 3.6 μm, 0.937 at 4.5 μm, 0.772 at 5.8 μm and 0.737 at 8.0 μm. However, the precise value of this correction also depends on the underlying surface brightness distribution of the source, and for objects with size ~several arcminutes, such as is the case for the present source, it is counselled to use corrections which are somewhat smaller (i.e. between the values cited above and unity). The handbook concludes that "this remains one of the largest outstanding calibration problems of IRAC".

We have, in the face of these problems, chosen to leave the flux ratio mapping unchanged. The maximum correction factors for the 8.0μm/4.5μm and 5.8μm/4.5μm ratios are likely to be > 0.8, but less than unity, and ignoring this correction has little effect upon our interpretation of the results. By contrast, we have used the maximum correction factors for our large aperture photometry, implying that our 5.8 and 8.0 μm fluxes may be somewhat under-estimated. Here again however, such uncertainties are not of too great a concern when interpreting the present sources, since spectral variations are considerably greater than any likely correction to the fluxes.

Taken all in all, it would therefore appear that some caution is required in interpreting the results of extended source photometry, as well as in determining and mapping the IRAC band ratios, although our present results are unlikely to mislead us in any serious way.

In addition to the GLIMPSE results, the inner 258 square degree of the Galactic plane has been surveyed through the MIPSGAL project. Mapping was obtained at 24 μm and 70 μm using the Multi-band Imaging Photometer for Spitzer (MIPS), the seminal paper for which is provided by Rieke et al. (2004). The 24 μm pixel sizes are 1.25 arcsec, and the spatial resolution of the results is 6 arcsec. Here again, we provide 24 μm mapping of the wind-swept globule G357.46+0.60, as well as presenting larger scale 0.54 x 0.54 deg$^2$ images of the region of interaction between the SNR and interstellar medium (ISM). Photometry of the YSO at $l$ = 357.4607°, b = 0.5988° was performed through small-aperture integration of the FITS results, where care was again taken to evaluate background emission at locations close to the source.

The 2MASS all-sky survey was undertaken between 1997 and 2001 using 1.3 m telescopes based at Mt Hopkins, Arizona, and at the CTIO in Chile. We have used results from the PSC to evaluate upper limit flux values for the YSO in G357.46+0.60, and constrain the Spectral Energy Distribution (SED) of the source. These results were later used to limit the range of YSO model SEDs evaluated by Robitaille et al. (2006), and place constraints upon the luminosity and mass of the YSO.

IRAS was launched in January 1983, and tasked with mapping 96% of the sky in wavebands at 12 $\mu$m, 25 $\mu$m, 60 $\mu$m and 100 $\mu$m. The resolution ranged from 30 arcsec at 12 $\mu$m through to 2 arcmins at 100 $\mu$m. The cometary structure G357.46+0.60 was detected during this survey, and is listed as IRAS17338-3044 in the PSC. These fluxes are used to define the variation in the FIR continuum, and constrain the properties and temperatures of the dust.

Finally, various of our sources were observed using the SPIRIT III instrument on board the MSX, and this permitted observations at 8.28 $\mu$m, 12.13 $\mu$m, 14.65 $\mu$m and 21.3 $\mu$m. Photometry from this catalogue was used to define the spectrum of cloud G357.46+0.60.

## 3. The Region of Interaction between SNR G357.7+0.3 and the ISM

The SNR G357.7+0.3 was first discovered by Reich & Fürst (1984), who provided mapping of the source in the 6 and 11 cm radio continuum. These authors showed the source to have a roughly circular morphology with enhanced rim structure, typical of what would be expected for a hollow-shelled outflow with diameter ~24 arcmin. Further mapping at 834 MHz (Gray 1994) was also found to be consistent with that of Reich & Fürst (1984), whilst emphasising some irregularity of the outflow structure. Features noted in this mapping included a strong extension to lower Galactic latitudes (i.e. towards larger right ascensions and higher declinations), and some strengthening of emission to one side of the primary shell. It was suggested that the shell may be interacting with the local ISM, and that this was causing deformation of the structure.

Further evidence for interaction with the ISM was also observed by Yusef-Zadeh et al. (1999), who found evidence for OH(1720 MHz)

emission along a segment of the SNR rim – a region of space which we shall also find, below, to be associated with wind-disrupted globules. Such emission appears to be characteristic of regions which are being shocked by SNRs (see Sect. 1).

A superposition of the 11 cm map of Reich & Fürst (1984) upon GLIMPSE II 8 $\mu$m imaging of the region is illustrated in Fig. 1. It can be seen from this that there is a bright-rimmed cometary structure located towards the upper right-hand limits of the shell, with tail pointing radially away from the centre of the SNR. The cometary structure also lies within the range of OH 1720 MHz emission detected by Yusef-Zadeh et al. (1999). This structure, however, is only a part of a more extended region of emission, in which a broad range of clouds are found to possess wind-swept morphologies.

Thus for example, we show a MIPSGAL 24 $\mu$m image of this region in Fig. 2, centred upon $\ell$ = 357.23°, b = 0.785°. The cometary structure referred to above is located to the lower left-hand size of the frame, and is seen to represent one of a broad range of clouds which are similarly aligned. We shall be suggesting that all of these structures are phenomenologically related, and that all of them were created through interaction with the SNR shell.

It is of interest, with regard to these latter structures, to note that although their relationship with the SNR appears hardly to be in doubt, much of the region lies just beyond the outermost contours of the 11 cm mapping. This may imply that the shell is somewhat larger than deduced by Reich & Fürst (1984) and Gray (1994) – a scenario which is by no means implausible, given that these portions of their maps are undersampled and/or relatively weak – or that the region has been scoured by strong pre-cursor winds, related to the progenitor of the supernova and/or associated cluster.

An analysis of the MIR colours of stars within this region, undertaken using the GLIMPSE II PSC, shows that several sources have [3.6]-[4.5] and [5.8]-[8.0] indices in a range characteristic of Class I YSOs. These sources are indicated by blue disks in Fig. 3, where they are compared with trends determined from the YSO modelling of Allen et al. (2004). The sources are also identified using white circles within Fig. 2 (where it should be noted that a couple of the sources are located very close to one another, and are therefore represented by a single circle within Fig. 2). This listing of likely Class I YSOs probably represents a severe undersampling, however, and simply corresponds to cases for which significant fluxes were detected in all four of the IRAC bands. Many more sources are present in which [5.8]-[8.0] or [3.6]-[4.5] indices are large, but which remain undetected in one or more of the IRAC photometric channels.

It is clear that several these sources are also associated with wind-blown structures, and that five or six are located close to the region at $\ell$ = 357° 04', b = 0° 34.5'.

One of the best defined of these structures appears to be located at $\ell$ = 357° 27' 55", b = 00° 36' 14", however, and this is also the most closely located to the edge of the radio shell. This source has been detected in the FIR (by IRAS and MSX), and also observed at millimetric wavelengths (Fontani et al. 2005) and 2.695 GHz (Reich et al. 1984). We shall discuss it in considerably more detail in Sects. 4 & 5 below.

## 4. The Cometary Structure G357.46+0.60

The structure of the cometary source G357.46+0.60 is illustrated in Fig. 4, where we show mapping at wavelengths ranging between 3.6 and 24 $\mu$m, taken from data products deriving from the GLIMPSE II survey and MIPSGAL project. We also show a combined colour image of the source in the lower right-hand panel, where blue corresponds to 3.6 $\mu$m, green to 4.5 $\mu$m, orange to 5.8 $\mu$m, and red corresponds to 8.0 $\mu$m.

It can be seen from these that the 3.6 and 4.5 $\mu$m emission, for which flux levels are closely comparable, is confined to the lower limit rim of the primary emission structure, and likely arises from a shock ionised region responsible for bremsstrahlung emission. This leads to the yellow colouration of the rim noted in the lower right-hand panel of Fig. 4. As one passes towards the results at 5.8 and 8.0 $\mu$m, however, then it is clear that emission becomes spatially more extended, and that overall flux levels increase dramatically. This latter tendency is also noted in the spectrum in Fig. 5; where the filled circles correspond to emission from the cometary structure, and open squares to a probable

YSO. This latter source is also identified with a green circle in Fig. 4 (lower right-hand panel). We have finally included 2.695 GHz radio continuum measures due to Reich et al. (1984), taken using a 3 channel receiver mounted on the 100 m Effelsberg Telescope; a facility which is operated by the Max Planck Institut für Radioastronomie in Bonn, Germany. Similarly, the 1.2 mm fluxes are taken from Fontani et al. (2005), acquired using the SIMBA bolometer array mounted on the 15 m Swedish-ESO Submillimetre Telescope (La Silla, Chile). Finally, the upper limit NIR fluxes are derived from the 2MASS survey (see Sect. 2).

The peaking of the fluxes at 8 $\mu$m, and the sharp increase in emission between 4.5 and 5.8 $\mu$m, may be indicative of 6.2 $\mu$m PAH emission bands within the 5.8 photometric channel, and of the 7.7 and 8.6 $\mu$m PAH features in the 8.0 photometric band; features which are also commonly observed in HII regions and planetary nebulae (e.g. Peeters et al. 2002; Hony et al. 2001; Phillips & Ramos-Larios 2008a, b, c). Longer wave MIPS and IRAS fluxes (at 12 $\mu$m, 24 $\mu$m, 25 $\mu$m, 60 $\mu$m and 100 $\mu$m), however, appear to indicate the presence of a broader dust continuum.

Finally, the structure at 24 $\mu$m is similar to that observed at 5.8 and 8.0 $\mu$m (see Fig. 4), although the point source at $\ell$ = 357.4607°, b = 0.5988° appears to be relatively more important, and dominates the appearance of the rim.

The structure of the rim is finally shown in clearer detail in Figs. 6, where we show scans across the rim of G357.46+0.60 in all four of the IRAC bands, and for two directions of traverse. In the upper panel, the slice is taken along the approximate major axis of the structure, and reveals that the rim is up to ~30 times stronger in 8 $\mu$m than it appears to be at 3.6 or 4.5 $\mu$m.

Peak values of emission occur within 11 arcsec of the front of rim, after which fluxes decline more slowly towards negative relative positions, in a manner similar to that expected for cooling within a post-shock dissociative regime.

Similar trends are also shown in the lower panel in Fig. 6, where we illustrate logarithmic profiles across the point source feature. These results suggest that the latter source is spatially unresolved in all of the spectral bands. They also confirm that the 3.6 and 4.6 $\mu$m emission is confined to within 22 arcsec of the star, although the 5.8 and 8.0 $\mu$m emission appears to be much more extended.

Finally, we note that 5.8$\mu$m/4.5$\mu$m and 8.0$\mu$m/4.5$\mu$m ratios are rather similar, and peak at similar locations across the source. They are also, in many cases, extremely variable with position. Given that the 4.5 $\mu$m channel contains no PAH emission features, and those at 5.8 and 8.0 $\mu$m are often dominated by such bands (see our comments above), then it is conceivable that we may be observing very strong variations in the strengths of the PAH emission features.

## 5. The Spectral Characteristics of the Cometary Structure G357.46+00.60

The spectrum of G357.46+00.60 is illustrated in Fig. 5, and extends from 3.6 $\mu$m fluxes deriving from the present IRAC results, through to the 2.695 GHz measures of Reich et al. (1984). There are several characteristics of this spectrum which enable us to constrain the physical properties of the source. In the first place, the lower wavelength results at 3.6 and 4.5 $\mu$m clearly imply closely similar levels of flux; emission which is solely confined to a marginal sliver of the source, and arcs around the front face of the globule, as noted in the contour mapping illustrated in Fig. 4. Such emission presumably arises from shock ionisation due to interaction with the SNR.

There are various possible causes for the fluxes within these bands. They may for instance arise from 3.3 $\mu$m PAH band feature within the 3.6 $\mu$m channel, and continuum emission from grains – although the temperatures of the dust, for this latter case, must necessarily be of order >800 K, and this may require the stochastic heating of very small grains (see e.g. Draine 2003). The viability of such a mechanism would depend upon whether local radiation field energy densities are sufficiently high, and whether the processes of small grain destruction and formation are appropriately equilibrated: whether for instance the shock destruction of very small grains is more than compensated by the post-shock shattering of larger grains, and creation of smaller dust particles (see e.g. Allain, Leach & Sedlmayr 1996; Jones, Tielens & Hollenbach 1996).

An analysis of the MIR spectrum of IC 443, another region where a supernova remnant seems to be colliding with interstellar gas, also points to the likely presence of shock enhanced transitions, particularly the S(0) though S(7) rotational transitions of $H_2$ (Neufeld & Yuan 2008). Such transitions, where they occur, may dominate fluxes in all of the IRAC wavebands.

Finally, we note that fluxes are very closely similar in both the 3.5 and 4.5 µm channels, and this is also what would be expected where emission arises from shock compressed and ionised gas – plasmas for which emission measures may be appreciable, and levels of free-free and bound-free emission are substantial.

Given this latter possibility, and assuming the gas to be optically thin, then fluxes would be expected to increase to longer wavelengths, and become optically thick at "turn-over" wavelengths $\lambda_{TO}$ < 590 µm. By contrast, the radio continuum measured at 2.695 GHz ($\equiv \lambda$ = 1.11 $10^5$ µm) is only a little less strong than the millimetric flux of Fontani et al. (2005), and it is therefore likely that this latter emission is strongly affected by the extrapolated radio free-free continuum. Where this is the case, and one supposes that the continuum at 2.695 GHz is optically thin, then this would imply a turn-over wavelength $\lambda_{TO}$ > 1.11 $10^5$ µm.

The two turn-over wavelengths for the plasma continuum spectrum, one relating to emission in the MIR, and the other to that observed in the radio, would therefore be markedly different, and suggest radically differing physical environments. If for instance one adopts the analysis of Olnon (1975) for the continuum spectra of a variety of plasma geometries, then it is determined that spectral turn-over frequencies $v_{TO}$ can be represented through

$$\log\left(\frac{v_{TO}}{GHz}\right) = -0.516 + \frac{1}{2.1}\log\left[\left(\frac{T_e}{K}\right)^{-1.35}\left(\frac{E(0)}{cm^{-6}\,pc}\right)\right] \quad\ldots\ldots\ldots(1)$$

The various cases considered by Olnon (1975), which include spherical and cylindrical configurations, and uniform, power-law and Gaussian variations in electron density, are only barely applicable (if at all) to the case in hand. Nevertheless, given that the parameter E(0) corresponds to emission measure through the centre of the source (e.g. is given by $E(0) = 2n_e^2 R$ for an homogenous spherical envelope with radius R), then it is probably fair to assume that estimates of E(0) derived from our present turn-over frequencies are related to emission measures through the cometary structure.

Given that this is the case, and that electron temperatures $T_e \sim 10^4$ K, it follows that the values $E(0) > 1.5 \; 10^{12}$ cm$^{-6}$ pc determined for the 3.6 and 4.5 $\mu$m plasma emission, and $E(0) < 2.5 \; 10^7$ cm$^{-6}$ pc derived for the radio continuum component of the continuum, imply differences in the emission measures by a factor of $> 6 \; 10^4$.

Given now that the 3.6 and 4.5 $\mu$m emission is confined to depths comparable to the size of the rim in Fig. 4, and that the distance to the structure is 4.0 kpc (see our further discussion below), then this may indicate that this particular component of emission is associated with densities $n_e > 1.5 \; 10^6$ cm$^{-3}$. By contrast, where the lower frequency continuum measures of Reich et al. (1984) arise from a broader region of the outflow, comparable in size to the head of the structure illustrated in Fig. 4 (~ 1.4 arcmin), then the corresponding densities would be in the region of $n_e < 4 \; 10^3$ cm$^{-3}$.

Such an analysis therefore suggests that the ionised gas is associated with a broad range of densities, although our limits upon $n_e$ remain somewhat speculative at present. Such a result would however, should it be confirmed, hardly come as very much of a surprise. It is clear that large ranges in $n_e$ are likely to be characteristic of regions of this type.

As one advances to larger wavelengths between 5.8 and 12 $\mu$m, then it is clear that fluxes increase very rapidly indeed, and result in a secondary peak located close to 8 $\mu$m. This strongly suggests, as noted above, that we may be observing emission by the 5.8, 7.7 and 8.6 $\mu$m components of PAH band emission; components which may conceivably be excited by FUV photons generated by re-combination behind the rim shock structure.

Subsequently, the increase in fluxes to ~100 μm, defined though our present MIPS results, and photometry deriving from the IRAS and MSX missions, suggests the presence of a broadly peaked continuum associated with warm dust emission. It is probable that much of this emission derives from the globule G357.46+00.60, although it is also possible that at least some of the flux corresponds to an embedded YSO – a source which will be considered in further detail in our discussion below.

We have fitted these latter results using dust continuum functions $F_\nu \propto B_\nu \epsilon_\nu$, where $B_\nu$ is the Planck function, and $\epsilon_\nu \propto \nu^\beta$ corresponds to the emissivity of the dust. Where the grains are large, then one anticipates that $\beta \cong 0$. On the other hand, where dust grains are very small ($a/\lambda \ll 1$), , then it is more likely that exponents $\beta$ will take values closer to ~1-2, depending upon the composition of the dust.

It can be seen that the low values of millimetric flux appear to constrain $\beta$ to values of order ~ 2. Larger exponents ($\beta$ ~ 0 or 1) would yield fluxes which are orders of magnitude too high - a disparity which is made all the worse if one also includes the differing plasma contributions.

It is therefore clear that the overall spectrum of the source, only partially defined though it may be, nevertheless enables us to place useful constraints upon conditions within the source.

Given that we are almost certainly dealing with a pre-existing molecular cloud interacting with the SNR outflow G357.7+0.3, then it is pertinent to ask whether there may be any evidence for triggered star-formation, such as was considered in Sect. 1.

The answer, as we shall note below, is almost certainly in the affirmative, and it is likely that the bright unresolved condensation in G357.46+00.60 corresponds to a recently formed YSO.

## 6. Triggered Star Formation in the vicinity of G357.46+00.60

We have used the GLIMPSE II PSC to investigate the colours of point emission sources within a region of size 2x2 arcmin$^2$ centred on the source G357.46+00.60. The results of this analysis are illustrated in Fig. 3, where the sources are indicated using filled red symbols.

It is clear from this that most of the sources are located close to the regime of Class III YSOs and main sequence stars (red-filled disks) – that is, have colours similar to those of highly evolved YSOs, for which accretion disks make little if any contribution to the SEDs. The one exception to this is G357.461+0.599, which is indicated by the red rectangle to the right-hand side; an object which corresponds to the bright point source in the rim structure of G357.46+00.60. This appears to be the sole object located among the Class I YSOs.

In fact, this latter object is also one of the few sources to possess only 4.5, 5.8 and 8.0 $\mu$m fluxes; the 3.6 $\mu$m flux is not recorded within the PSC. This presumably arises because of the weakness of this source at shorter IRAC wavelengths, and the difficulty in distinguishing between this particular object and the broader, resolved components of emission. We were therefore required to estimate [3.6] magnitudes using the procedures described in Sect. 2, and these yielded colours [3.6]-[4.5] = 1.30, and [5.8]-[8.0] = 0.79.

This source is not evident in shorter wave 2MASS results, where its location is flanked by two sources in the J photometric band, each separated by 2 arcsec from the MIR position. These two (unrelated?) sources subsequently merge in H and $K_S$, perhaps as a result of increasing levels of emission associated with the intermediate MIR source. The fluxes in Figs. 5 & 7 correspond to realistic (and conservative) upper limits based upon these rather confusing trends.

Finally, we have also included, in this spectrum, the results of photometry of the 24 $\mu$m MIPS imaging results. This, in combination with all of the other photometry, appears to indicate that the object has a steeply increasing spectrum, quite untypical of those observed for other stars in this region. We provisionally identify it as a Class I YSO – and indeed, note that no normal reddened stellar spectrum appears capable of replicating these results.

It therefore appears that there is only one candidate source that can be identified as a YSO, that is located within 1 arcmin of G357.46+00.60, and that is brighter than the sensitivity limits of the PSC. This does not however exclude the possibility that there are many much fainter YSOs – sources corresponding to Class 0 structures (for which much of the emission occurs in the FIR), or to very much lower mass stars. It would be rather surprising, indeed, if this were not the case.

The mechanisms which gave rise to this star are not entirely easy to pin down, and it may represent a case in which pre-existing star formation is associated with later, and otherwise unrelated wind-blow structures. We believe that this is unlikely to be what is happening in this particular case, however. This and several of the other YSO candidates (see Fig. 2) appear to be perched at the leading edges of the interaction zones, where the bright-rimmed structures are observed to occur. These are precisely the regimes where one would have expected them to be located if their formation had been triggered by the supernova event. It is therefore plausible to suppose that many of these YSOs developed through interaction between the clouds and the SNR, followed by gravitational instability in the shock compressed gas.

Such a model also requires that the age of the YSO is comparable or less than that of the SNR. We shall consider this question, and the likely physical characteristics of the YSO, in the following analysis of this source.

## 7. The Characteristics of the YSO within the rim of G357.46+00.60

In analysing the characteristics of the YSO described in the previous section, we need to take note that the distance of the source is far from well established. Thus for instance, Leahy (1989) determines a distance of 6.4 kpc to the SNR using the $\Sigma$-D relation of Li & Wheeler (1984) – that is, the relation between the 1 GHz surface brightness $\Sigma$ and the distance to the SNR. Such a relation is almost next to useless for most such sources, however, and gives little more than an order of magnitude indication of the value of D. Green (2005) has noted that $\Sigma$-D trends are extremely uncertain, show a large degree of scatter, and are very poorly grounded in a theoretical understanding of SNR surface brightnesses.

The millimetric results of Fontani et al. (2005), by contrast, are shown to imply a distance of 4.0 pc or 13.0 pc, where they have used CS line radial velocities, and assumed a normal Galactic rotational curve. We shall, for consistency with the SNR distance, be using the former of these estimates in the following analysis. We shall also be assuming that this value has a higher degree of verisimilitude than that of the $\Sigma$-D result.

Even this value, however, is not entirely to be trusted, and may be subject to appreciable errors and uncertainties. Given that the dynamics of source G357.46+00.60 are strongly affected by the SNR, and that we may also be observing shocked (and accelerated) material within the main body of the source, then there is a possibility that CS J=2-1 and 3-2 line velocities, which normally refer to denser portions of the gas, may arise from recombining material in a post-shock dissociative regime. There is also the possibility that the globule underwent asymmetric acceleration in the radiation field of the original stellar cluster, a mechanism which was briefly described in Sect. 1.

All of these effects may lead to differences between the observed velocity of the cloud and that of the local standard of rest - and cause distances to be underestimated or overestimated, depending upon the precise location of the source.

Having said this, however, we note that that the $\Sigma$-D and millimetric analyses imply similar distances to the source. There is therefore hope that the millimetric results may be providing a tolerably reasonable estimate of D.

Given this value of distance, it is then possible to evaluate the type of YSO which we may be observing, and place constraints upon the luminosity and mass of the source.

To do this, we make use of the $\sim 2\ 10^5$ YSO models evaluated by Robitaille et al. (2006), analysed using the SED fitting tool developed by Robitaille et al. (2007), in which SEDs have been determined for differing dust and gas geometries, varying dust properties, and using 2D radiative transfer modelling for a large region of parameter space. The fitting of the models also takes account of the apertures employed for the photometry. Given that our present source does not appear to be resolved in any of the wavebands, we have set this parameter to be equal to the FWHM of the PSFs in each of the respective channels – ranging from ~2 arcsec in the IRAC wavebands to ~6 arcsec for the MIPS results. An example of the fitting of such models is provided in Fig. 7, where we have required that $\chi^2-\chi^2_{BEST} < 3$ (see Robitaille et al. (2006, 2007) for details).

The black curve corresponds to the best fitting model solution, for which the total $\chi^2$ is 4.74, whilst the lower dashed curve represents the corresponding stellar photospheric model. The grey curves are the solutions for the resolution limits cited above, whilst red SEDs are for an aperture of 6 arcsec, and blue curves for an aperture of 2 arcsec.

The resulting YSO properties for the best fit models are summarised in Fig. 8, where we represent the stellar luminosity, temperature and mass, together with a range of other parameters relating to the star. We also summarise some of the more important characteristics of the accretion disk and outer envelope. All of the parameters are represented against the lifetime of the source $T_{EV}$. Finally, the grey areas indicate the ranges parameter space occupied by the models, where each change in grey corresponds to a factor of 10 variation in model density, whilst the best fit models are indicated using small black bullets.

If one supposes, as we shall be doing here, that the formation of the YSO is triggered by shock interaction between a neutral cloud and the SNR, and that the SNR has a lifetime $\sim 10^4 (D/5\ kpc)^2$ yr (Leahy 1989), corresponding to $\sim 6.4 10^3$ years for the distance adopted here, then it is clear that the luminosity must be of order $\sim 5\ 10^2 \rightarrow 2\ 10^3\ L_\odot$, temperatures are $\sim 3000$ K, and that the stellar mass is likely to be between 5 and 10 $M_\odot$. We therefore appear, given that our input parameters are appropriate, to be dealing with the formation of a reasonably high mass YSO.

Apart from this, the present stellar radius is noted to be $\sim 30$-$70\ R_\odot$; the accretion disk mass is $\sim 0.05 \rightarrow 0.2\ M_\odot$; disk accretion rates may be anywhere between $10^{-7}$ and $10^{-4}\ M_\odot\ yr^{-1}$; and the disk radius is likely to be between 3 and 10 AU. This disk is, in turn, contained within a very much larger envelope with outer radius $\sim 10^4$-$5\ 10^4$ AU, and accretion rates of between $3\ 10^{-5}$ and $2\ 10^{-3}\ M_\odot\ yr^{-1}$. This large value for the envelope accretion rate would imply that the star is in a very early phase of formation; what Robitaille et al. (2006) refer to as Stage 0/I objects. Such objects are similar to (and in most cases probably identical with) those classified as Class 0 and I YSOs. The present results are therefore consistent with the placement of the source within Fig 3.

Where one therefore assumes that the formation of the YSO is triggered by SNR G357.7+0.3, then this permits one to tie down the characteristics of the source to moderate levels of precision. It is also interesting to note that one of the two ranges of $T_{EV}$ permitted by our current models corresponds to precisely the range that would be expected for triggered star formation within the source.

It is pertinent, given these results, to note that the cloud has been identified as a possible high mass star formation centre on the basis of its IRAS colours (Fontani et al. 2005 and references therein); an identification which persuaded Fontani et al. to investigate this and similar regions at millimetric wavelengths. In addition to the continuum fluxes employed in our analysis above, these authors also determine CS J = 2-1 and J = 3-2 line widths of ~ 2.4→2.5 km s$^{-1}$, and a diameter of 19 arcsec for the millimetric source. This latter diameter, in combination with their continuum flux, and the distance employed in our present work, is then used to determine an overall clump mass of 175 $M_\odot$; an estimate which is comparable to the value ~240 $M_\odot$ required for virial equilibrium. Although this agreement between the deduced and virial masses may seem to be persuasive, much depends upon whether CS velocities are due to motion within a gravitational potential well; the analysis presented earlier in this paper suggests that this may not be the case.

Finally, we note that this particular star, and the structure within which it is embedded, represent just one of several wind-blown structures within this regime. We take the present cloud, G357.46+0.60, to be reasonably typical of these other sources as well. However, there may be one exception to this rule represented by a conical emission structure in Fig. 2, located towards the upper right-hand side at $\ell$ = 357.14°, b = 0.84°. We shall briefly consider this source in the following section.

## 8. V1139 Sco – A Case of Interaction between a Mira Envelope and SNR?

Perusal of the field of wind-blown structures in Fig. 2 draws attention to an interesting source which is further illustrated in Fig. 9. This image consists of 3.6 $\mu$m→8 $\mu$m IRAC images (blue-to-orange colours), combined with a 24 $\mu$m image deriving from the MIPSGAL project (red). It would appear from this that a very bright star, identifiable as V1139 Sco, is located at the apex of a conical structure extending > 0.2° to the Galactic NW.

V1139 Sco has been identified as a Mira star with period 438.6 days (Alard et al. 1996), and possesses an R band variation of between 17.37 and 14.55 mags (Terzan et al. 1982). Combined photometry from the 2MASS, IRAS and MSX survey suggests that the continuum peaks close to the $K_S$ band (2.17 $\mu$m) or perhaps at very slightly longer wavelengths, and has a secondary emission peak close to ~12 $\mu$m. Given the quoted bandwidths of the MSX 12.13 $\mu$m filter in particular (2.1 $\mu$m at 50% peak intensity), and the relative intensity of this band compared to measures at 8.28 $\mu$m, then it seems probable that we are observing dust mission features arising from SiC grains.

The association of such a star with the wind-swept funnel must be counted as very unusual indeed, and might tempt one to suppose that the Mira wind is being swept backwards behind the star. If this is so, then it also follows that the SNR and Mira distances must be comparable.

To gain some handle on the distance to the star, we note that the 2MASS colours of V1139 Sco are given by J-H = 1.468 mag, and H-$K_S$ = 0.751 mag, whilst the $K_S$ band magnitude is 3.345 ± 0.266 mag. The error in this latter parameter is much greater than would be expected for a source of this brightness, and testifies to the variability of the source between the differing epochs of observation.

These latter values convert to J-H = 1.544, H-K = 0.726 mag and K = 3.389 when one corrects to the homogenized system of Bessell & Brett (1988) using the colour transformations of Carpenter (2001). Given that the star has been assigned a spectral type of M8 (Roharto et al. 1984), and that the intrinsic colours of such a star are close to J-H $\cong$ 0.96 mag,

H-K ≅ 0.31 mag (Bessell & Brett 1988), then this implies reddening excesses of order E(H-K) ≅ 0.42 mag and E(J-H) ≅ 0.58 mag.

Where one now employs the NIR extinction coefficients of Cardelli et al. (1989) for $R_V = A_V/E_{B-V}$ = 3.1, then this would imply the mean extinctions $A_V$ ≅ 5.82 mag, and $A_K$ ≅ 0.63 mag.

Using the light-curve period cited above, and employing the period-luminosity (P-L) relation of Groenewegen & Whitelock (1996) then this implies an average K band absolute magnitude $M_K$ = -8.18 mag; where we have also corrected the P-L relation for a more recent distance to the Large Magellanic Cloud (LMC) (Macri et al. 2006). Finally, combining the colour corrected K band magnitude derived from the 2MASS results, the mean absolute magnitude $M_K$, and the extinction estimate $A_K$, then one derives a distance to the star of 1.54 kpc.

The latter value assumes, however, that the 2MASS $K_S$ magnitude corresponds to the mean (period averaged) brightness of the source. Such is not, however, necessarily the case. If one assumes that the 2MASS $K_S$ magnitude corresponds to the peak source brightness in this waveband, and notes that the amplitude of Mira K band variability is typically of order ΔK = 0.4 to 1.2 mag (see e.g. Glass et al. 1990), then it follows that the distance to the star may be as great as 2.0 kpc.

There are various imponderables in this type of analysis. It is worth noting for instance that the period-luminosity relation of Groenewegen & Whitelock (1996) and others (e.g. Feast et al. 1989) are normally derived using LMC variables – stars whose metallicity is lower than that of their Galactic counterparts. We have only the vaguest idea of how this difference in metallicity may affect the luminosities of the stars. Nevertheless, it seems clear that the distance to V1139 Sco must be much less than that determined from the observations of Fontani et al. (2005).

This leads us, in turn, to two possible interpretations of the results. On the one hand, V1139 Sco may be much closer to us than the SNR, and therefore unrelated to the extended plume. Alternatively, it is possible that the distance to the SNR is much less than would be supposed from the radial velocity measures of Fontani et al. (2005), and perhaps of order ~half of the value employed in our analysis above.

This, if it were the case, would imply some changes to the YSO modelling, and suggest luminosities in the range L = 50-600 L$_\odot$ and masses between ~1 and 8 M$_\odot$; both of which ranges are lower than was determined above.

## 9. Conclusions

We have drawn attention to a region of wind-swept clouds in an extended region of sky centred at $\ell$ = 357.23°, b = 44.55°. These appear to be located near the limit of the radio shell associated with SNR G357.7+0.3, and are also associated, in certain cases, with the presence of OH (1720 MHz) maser emission. It is noted that the structures are radially directed towards the centre of the SNR, and likely affected by shock interaction with the SNR shell.

An analysis of sources within the GLIMPSE II PSC also suggests that Class I YSOs litter most of the region, with many of them appearing to be associated with the wind-blown structures.

One of the brightest of these clouds, and certainly the one that appears to be most closely located to the SNR radio shell, has a cometary morphology and bright rim structure. An analysis of this cloud shows evidence for strong variations in 8.0μm/4.5μm and 5.8μm/4.5μm ratios, perhaps attributable to small-scale variations in PAH emission intensities. We also note evidence for a broad dust continuum peaking close to λ ~ 100 μm. Given that this emission appears to fall to very low levels by λ = 1.2 mm, then this would imply an exponent of grain emissivity ~2, and associated grain temperatures T$_{GR}$ ≅ 30 K.

Finally, the 3.6 and 4.5 μm fluxes may be dominated by a variety of emission processes. Where bremsstrahlung emission is important, however, then this would imply emission measures >1.5 10$^{12}$ cm$^{-6}$ pc, and densities n$_e$ > 1.5 10$^6$ cm$^{-3}$. The radio continuum, by contrast, is associated with a much lower emission measure plasma (<2.5 10$^7$ cm$^{-6}$ pc), and perhaps implies electron densities in the range n$_e$ < 4 10$^3$ cm$^{-3}$.

The bright rim region of G357.46+0.60 also contains a bright, cool, and unresolved region of emission, with colours comparable to those of Class I YSOs. If this is presumed to have formed through a phase of recent star-formation, triggered by shock interaction between the cloud and SNR, then it would imply a mass $M_{YSO}$ ~ 5-10 $M_\odot$, and luminosity ~ $5 \; 10^2 \rightarrow 2 \; 10^3$ $L_\odot$.

Finally, we have drawn attention to a very interesting funnel-type structure associated with the Mira variable V1139 Sco. It is suggested that the Mira and the plume may be physically related, and that we are observing a situation in which the expanding Mira outflow is being swept backwards by the SNR. If this is the case, however, then the distance to the region must be $\leq$ half of that supposed from millimetric observations of G357.46+0.60, and imply lower luminosities and masses for the YSO.

**Acknowledgements**

We would like to thank the referee for his very interesting and useful comments on this paper, and acknowledge the help of Daisy Zepeda-Garcia in helping to contour certain of the maps. This work is based, in part, on observations made with the Spitzer Space Telescope, which is operated by the Jet Propulsion Laboratory, California Institute of Technology under a contract with NASA. Support for this work was provided by an award issued by JPL/Caltech. It also makes use of data products from the Two Micron All Sky Survey, which is a joint project of the University of Massachusetts and the Infrared Processing and Analysis Center/California Institute of Technology, funded by the National Aeronautics and Space Administration and the National Science Foundation. GRL acknowledges support from CONACyT (Mexico) grant 93172.

# Figure Captions

**Figure 1**

Superposition of 11 cm radio contours for SNR G357.7+0.3 (taken from Reich & Fürst 1984) upon an 8.0 $\mu$m image taken with the SST. Note the presence of a well-defined cometary structure to the upper right-hand side of the SNR; a feature which is likely to represent a case of shock interaction between a compact molecular cloud and the SNR.

**Figure 2**

A MIPSGAL 24 $\mu$m image of wind-blown structures located to the Galactic NW of SNR G357.7+0.3. Most of these point radially towards the centre of the SNR. The cometary structure to the lower left-hand side of the image is the same as the corresponding feature noted in Fig. 1, whilst the small white circles indicate the locations of probable Class I YSOs.

**Figure 3**

The regimes occupied by Class I, II and III YSOs within the MIR colour plane, based upon an analysis of YSO models presented by Allen et al. (2004). The blue disks correspond to the YSOs identified with white circles in Fig. 2. The red disks indicate sources located within 1 arcmin of the cometary globule G357.46+00.60. Finally, the source G357.48+00.60, indicated by the red rectangle, corresponds to the bright unresolved feature in the head of G357.46+0.60 (illustrated in Figs. 4 & 8).

**Figure 4**

Contour mapping and imaging for the cometary globule G357.46+0.60, where we show Spitzer IRAC results at 3.6, 4.5, 5.8 and 8.0 $\mu$m, MIPSGAL results at 24 $\mu$m, and a combined 3.6 $\mu$m (blue), 4.5 $\mu$m (green), 5.8 $\mu$m (orange), and 8.0 $\mu$m (red) image of the source in the lower right-hand panel. The area covered by this map is the same as that indicated by the white rectangle in Fig. 2. The contour parameters (A,B) are given by (7.5, 0.1080) at 3.6 $\mu$m, (7.0, 0.0718) at 4.5 $\mu$m, (8.0, 0.1579) at 5.8 μm, and (15.0, 0.1692) at 8.0 μm. The contour values at 24 μm are linear, with the lowest contour set at 4.9147 MJy sr$^{-1}$, and the contour interval given by 47.12 MJy sr$^{-1}$. Note the concentration of 3.6 and 4.5 μm emission within the bright rim structure, and the more widely distributed and stronger emission at longer MIR wavelengths. The position of a probable high-mass YSO is indicated by a green circle in the lower right-hand panel.

**Figure 5**

Spectrum of the cometary structure G357.46+0.60 (filled circles) and of the associated YSO (open squares), where we illustrate results deriving from 2MASS (upper limit YSO photometry); 3.6→8.0 μm IRAC band imagery; 24 μm MIPSGAL imaging; FIR photometry deriving from the IRAS and MSX point source catalogues; 1.2 mm fluxes published by Fonatani et al. (2005), and the 2.695 GHz fluxes determined by Reich et al. (1984). The curves are flagged with the parameters (β, $T_{GR}$), and correspond to differing values of the grain emissivity exponent β, and grain temperature $T_{GR}$.

**Figure 6**

Profiles though the head of G357.46+0.60 in the four IRAC photometric bands, where the directions of the profile are indicated in the inserted images, and the widths of the slices are in all cases 1.8 arcsec. The upper panel shows a slice along the major axis of the source, and reveals the steep increase in intensities near the head of the globule, and the slower fall-off in emission as one moves to negative relative positions. The lower panel, by contrast, shows the logarithmic variation through the presumptive YSO, whence it is apparent that the bremsstrahlung emission observed in the 3.6 and 4.5 μm bands is confined to within 22 arcsec of the YSO.

**Figure 7**

Fitting of YSO modeling results to the 2MASS and Spitzer photometry, where the 2MASS photometry corresponds to the upper limit results. The models derive from the analysis of Robitaille et al. (2006), using the SED fitting tool of Robitaille et al. (2007). The black curve represents the best model fit to our photometric results, whilst the lower dashed curve represents the photospheric continuum. Grey curves represent the best fit model SEDs for spatial resolutions of between 2 and 6 arcsec (see text for details), whilst blue curves represent these SEDs for resolutions of 2 arcsec, and red curves apply for resolutions of 6 arcsec.

**Figure 8**

The panels show the variation of YSO parameters for the model solutions illustrated in Fig. 7, plotted against the evolutionary age of the star $T_{EV}$. Grey areas represent the ranges of parameter space covered by the models, with each variation of tone indicating a factor 10 change in the density of model sampling. The black points correspond to the best-fit models identified in Fig. 7. Where the formation of the star is triggered by SNR G357.7+0.3, then it is likely that the age of the source will be $< 6.4 \; 10^3$ yrs.

**Figure 9**

An image of the conical outflow source centred on the Mira variable V1139 Sco, which corresponds to the bright star at the apex of the funnel (lower left-hand side of the image). The image is a combination of five separate sets of results, including IRAC images at $3.6 \rightarrow 8.0 \; \mu m$ (represented as blue-orange), and a 24 $\mu m$ image deriving from the MIPSGAL project (represented as red).

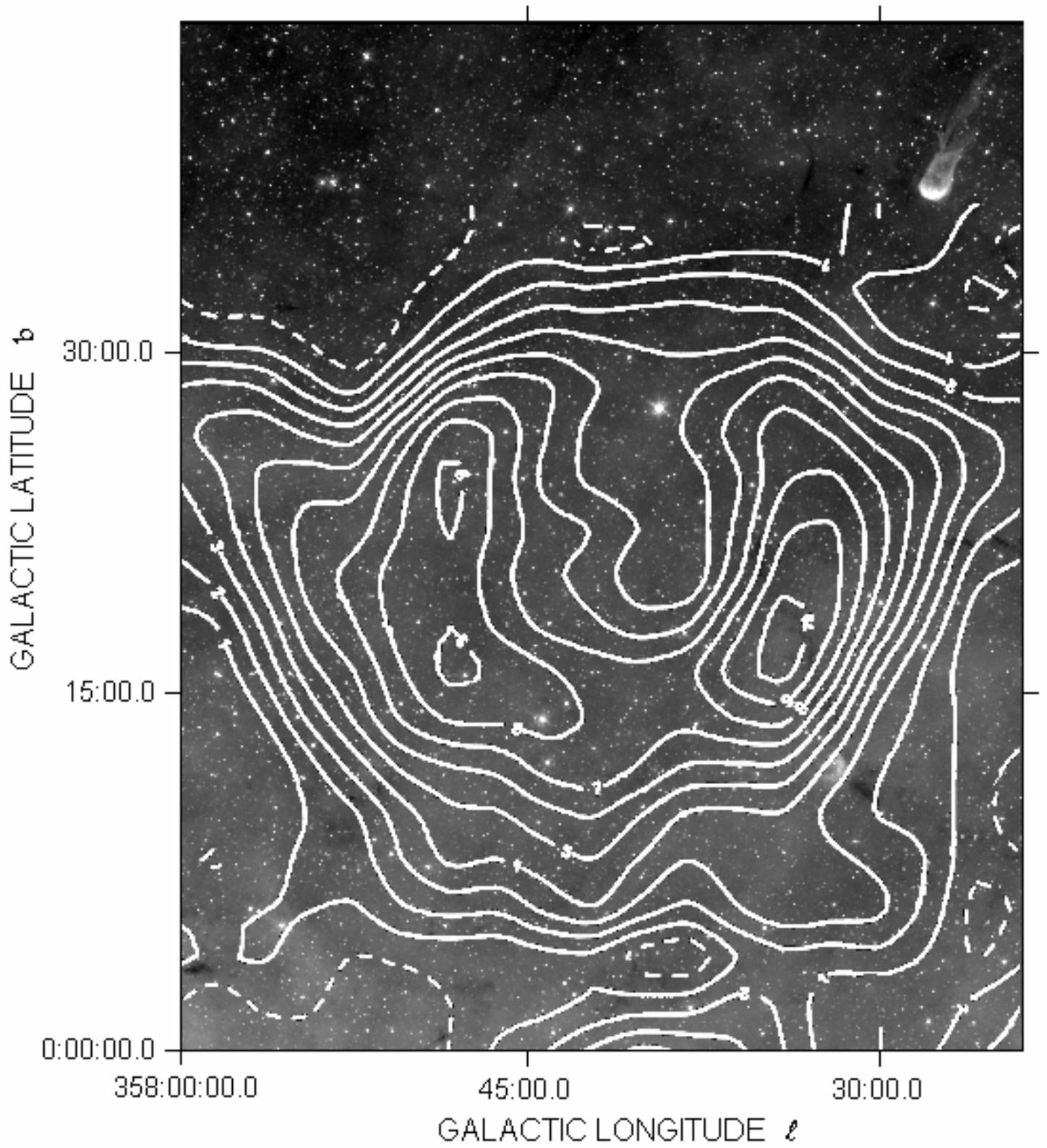

Figure 1

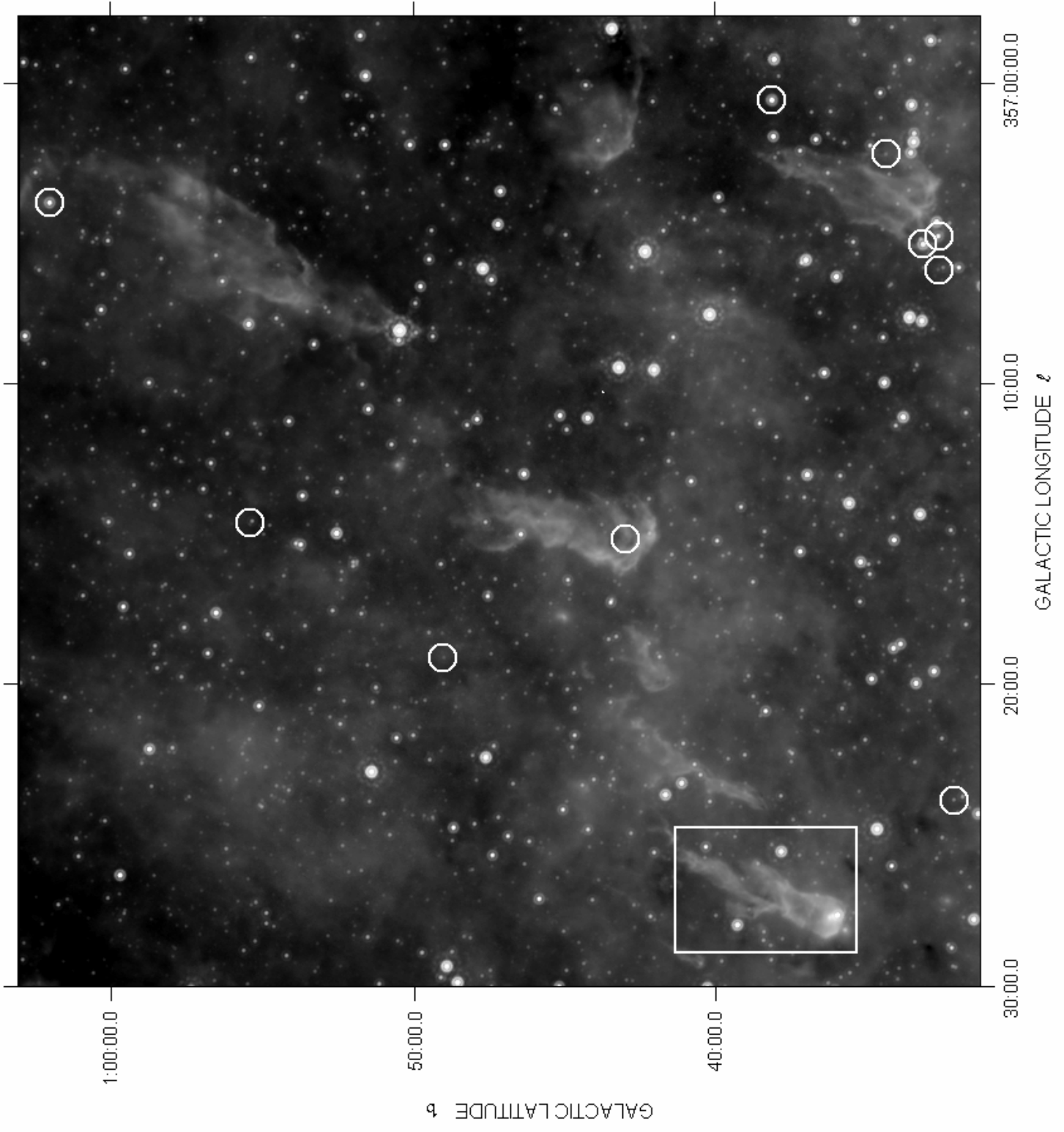

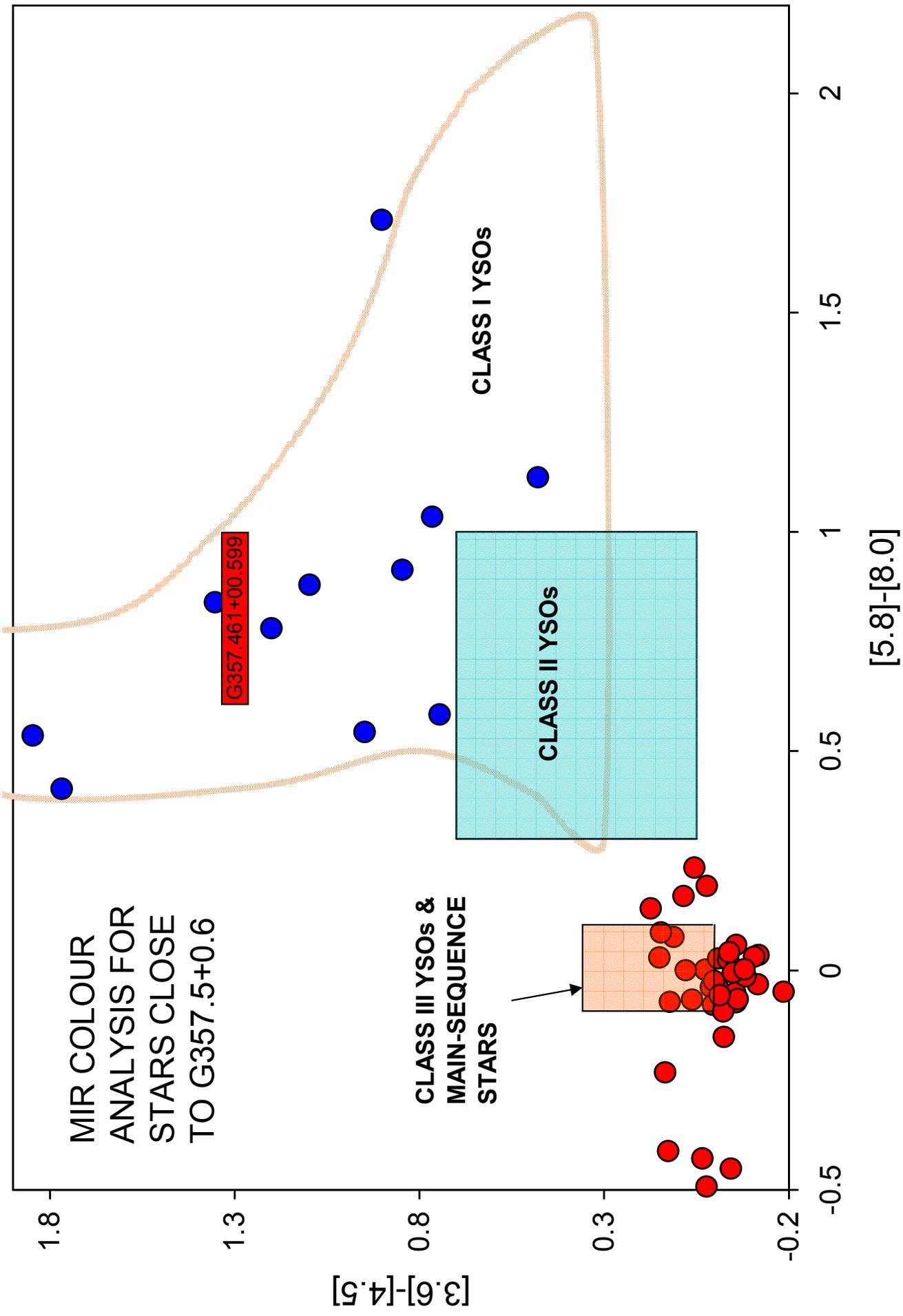

Figure 3

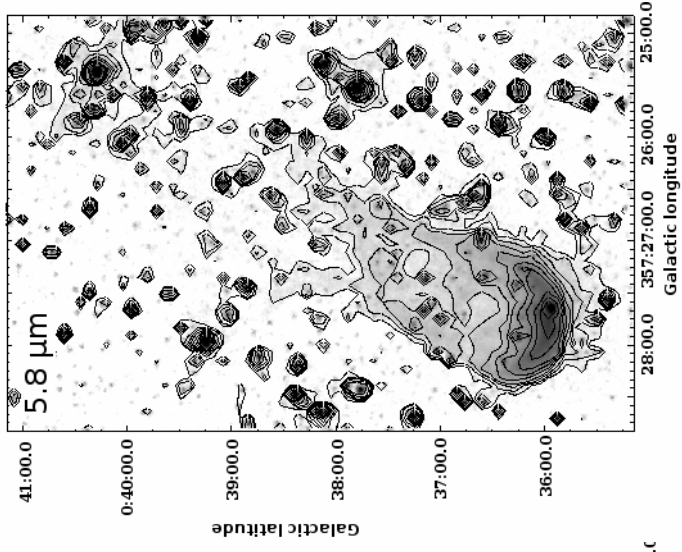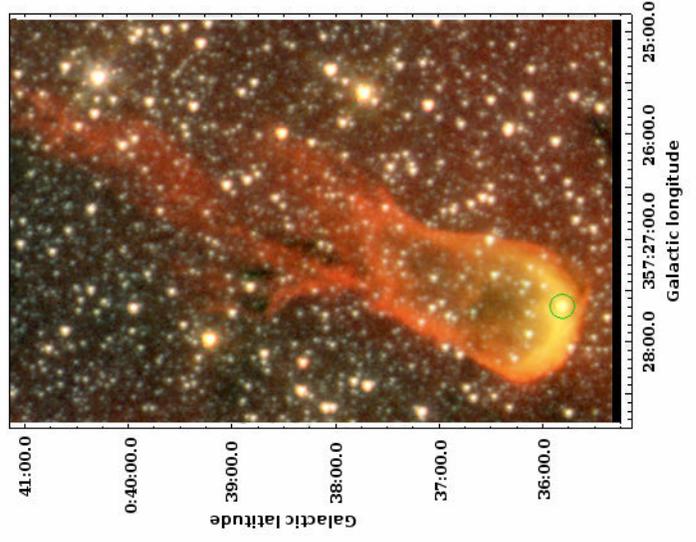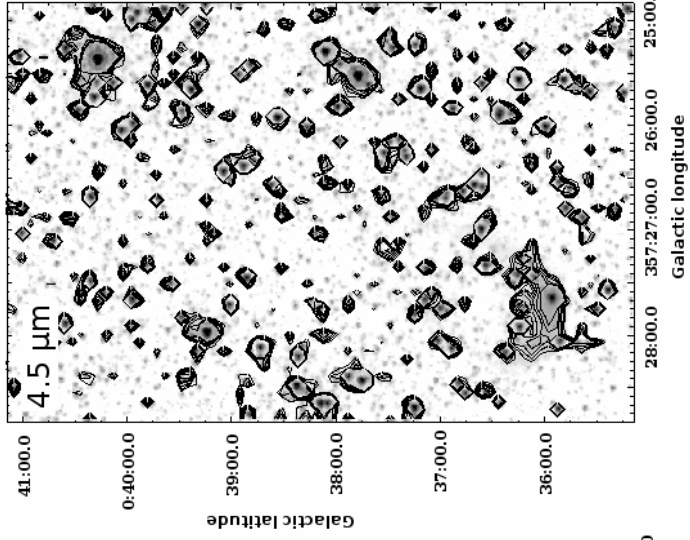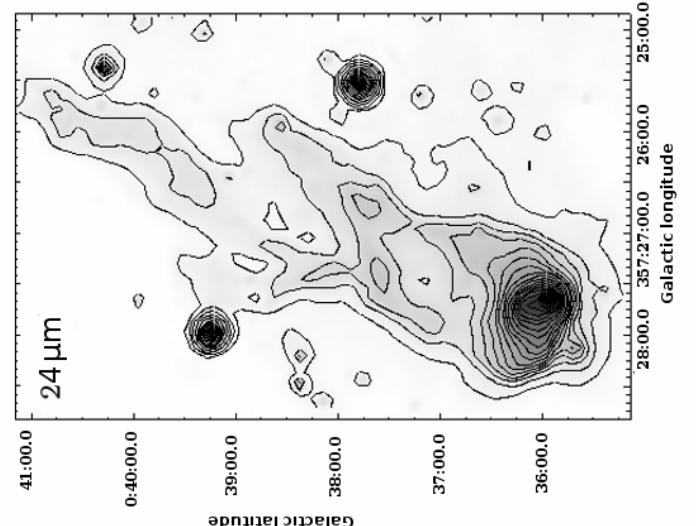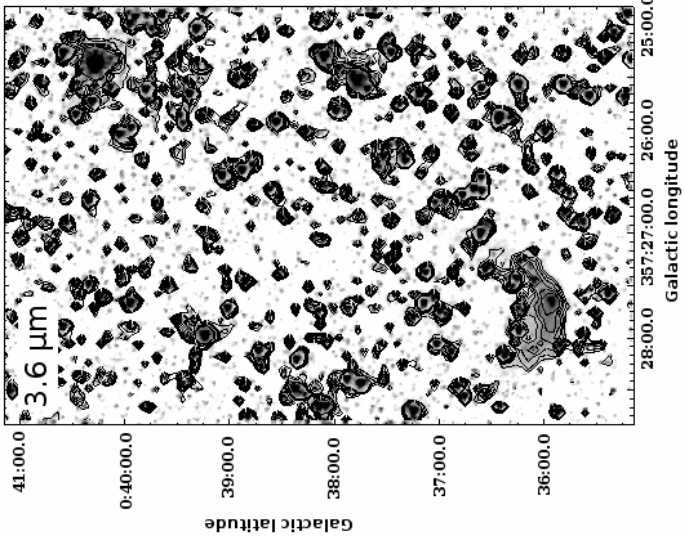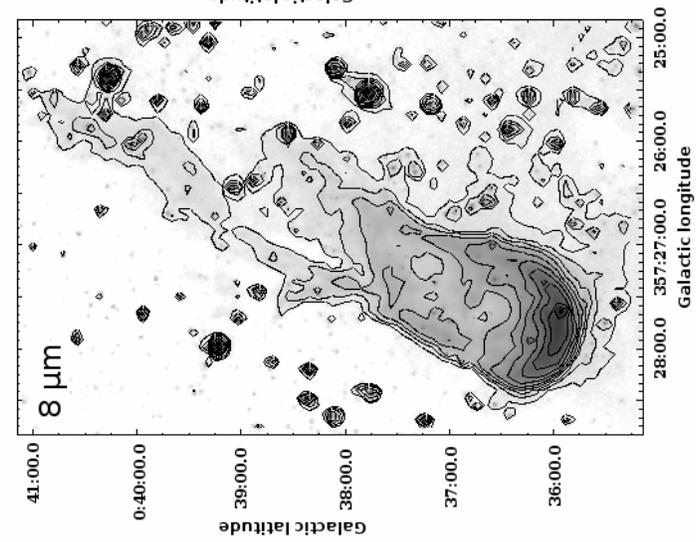

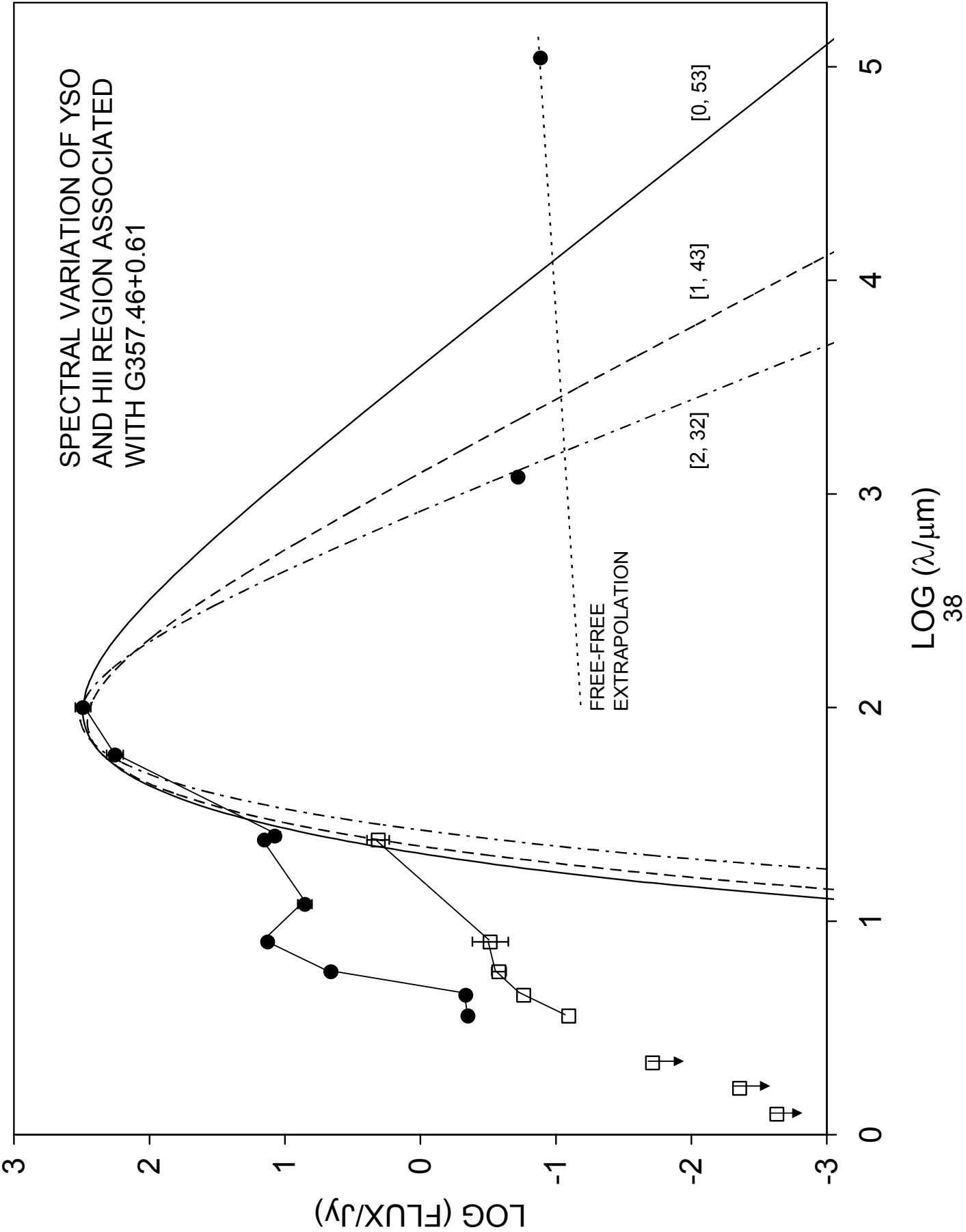

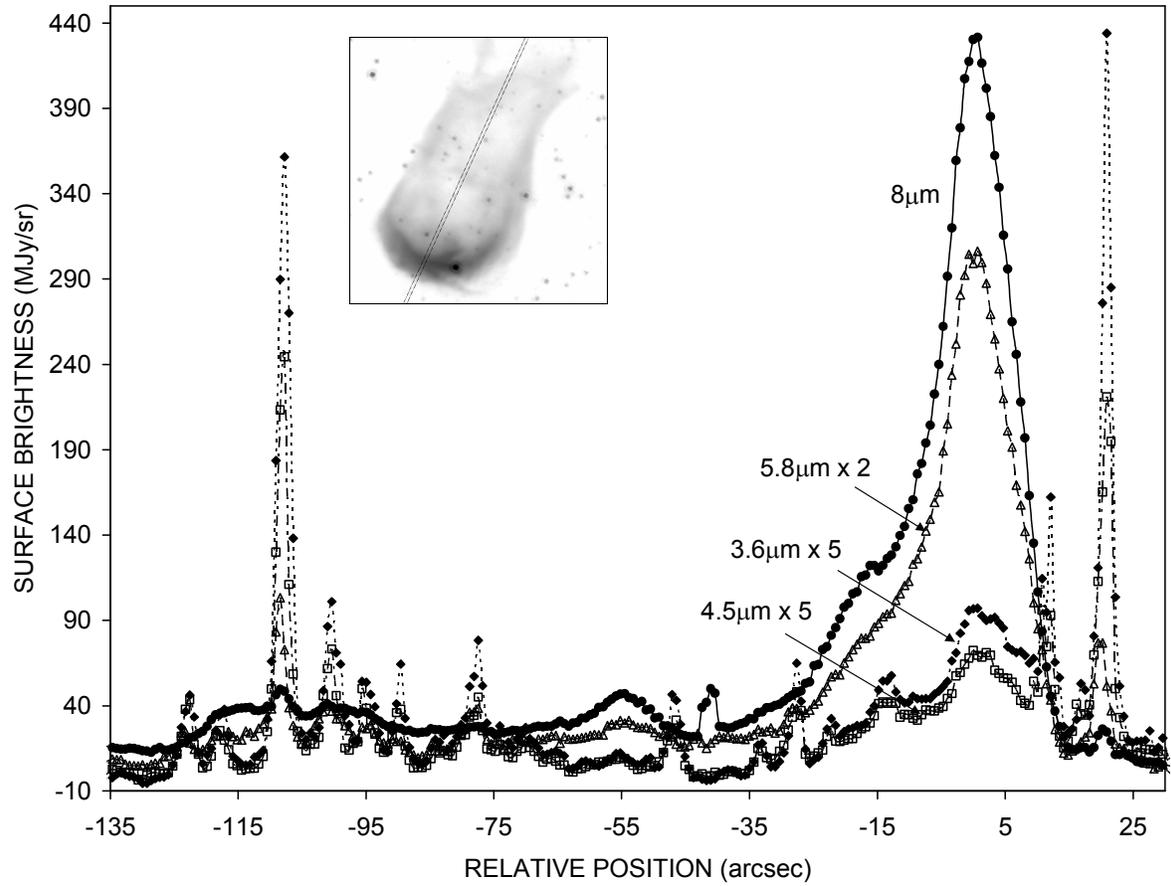
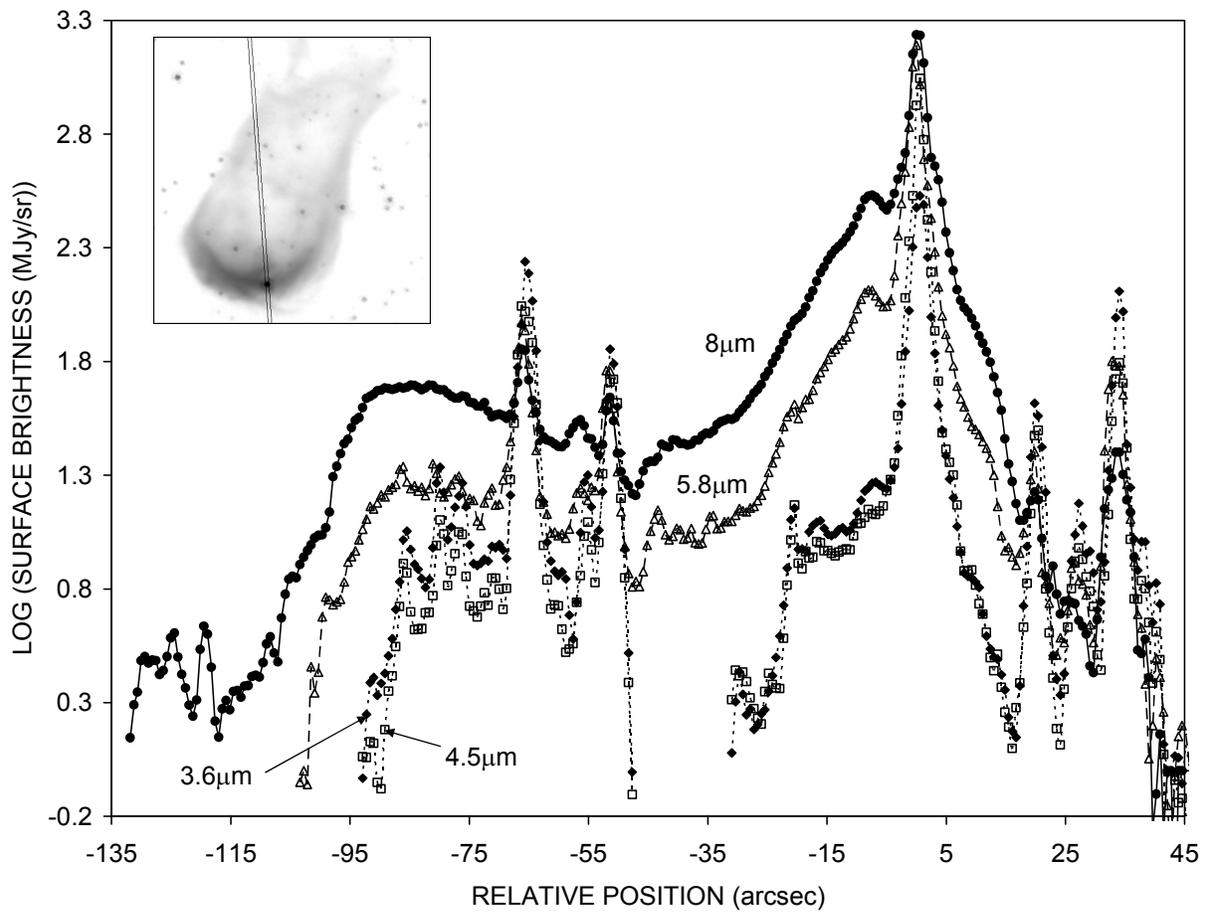

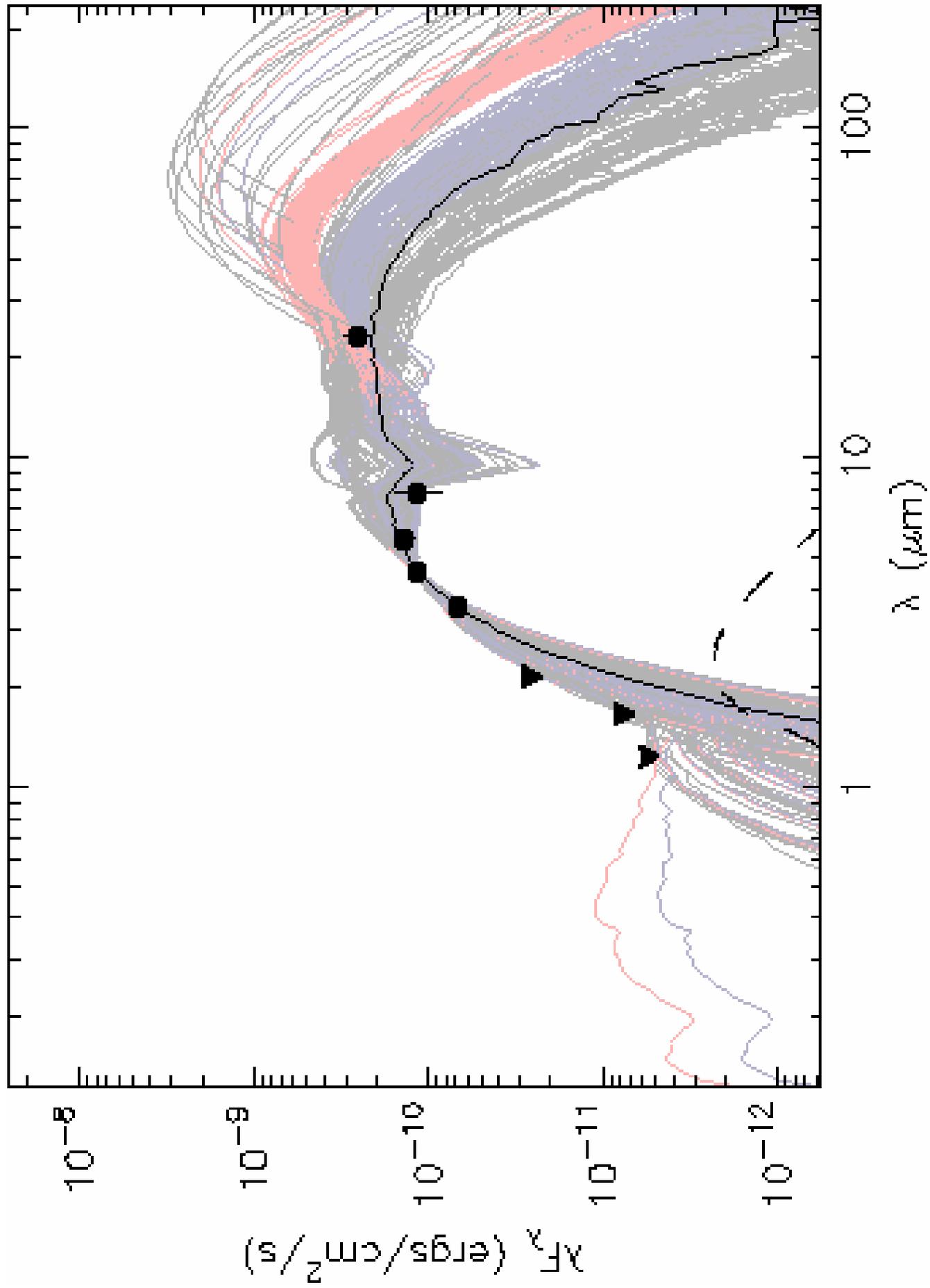

Figure 7

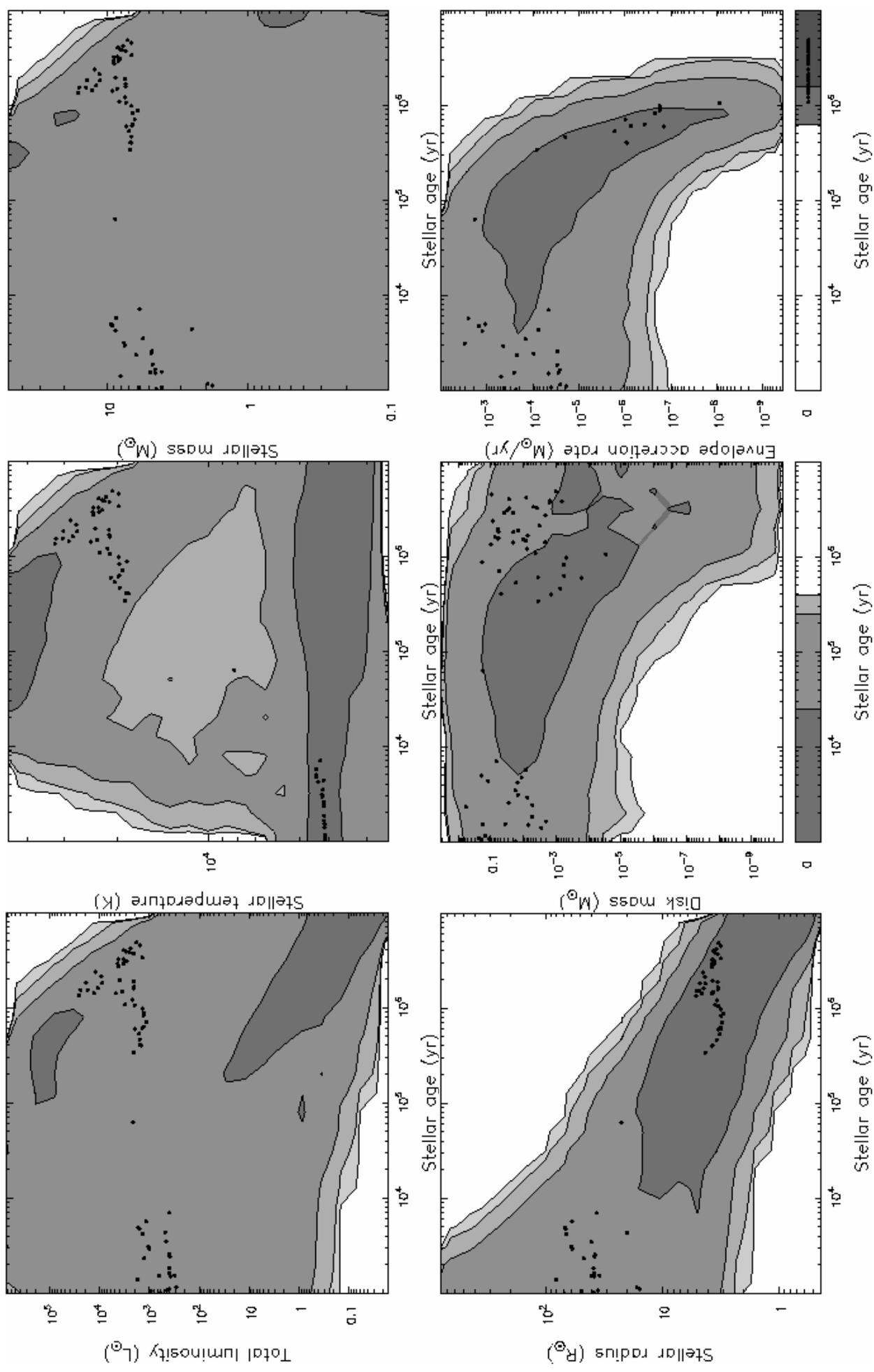

Figure 8

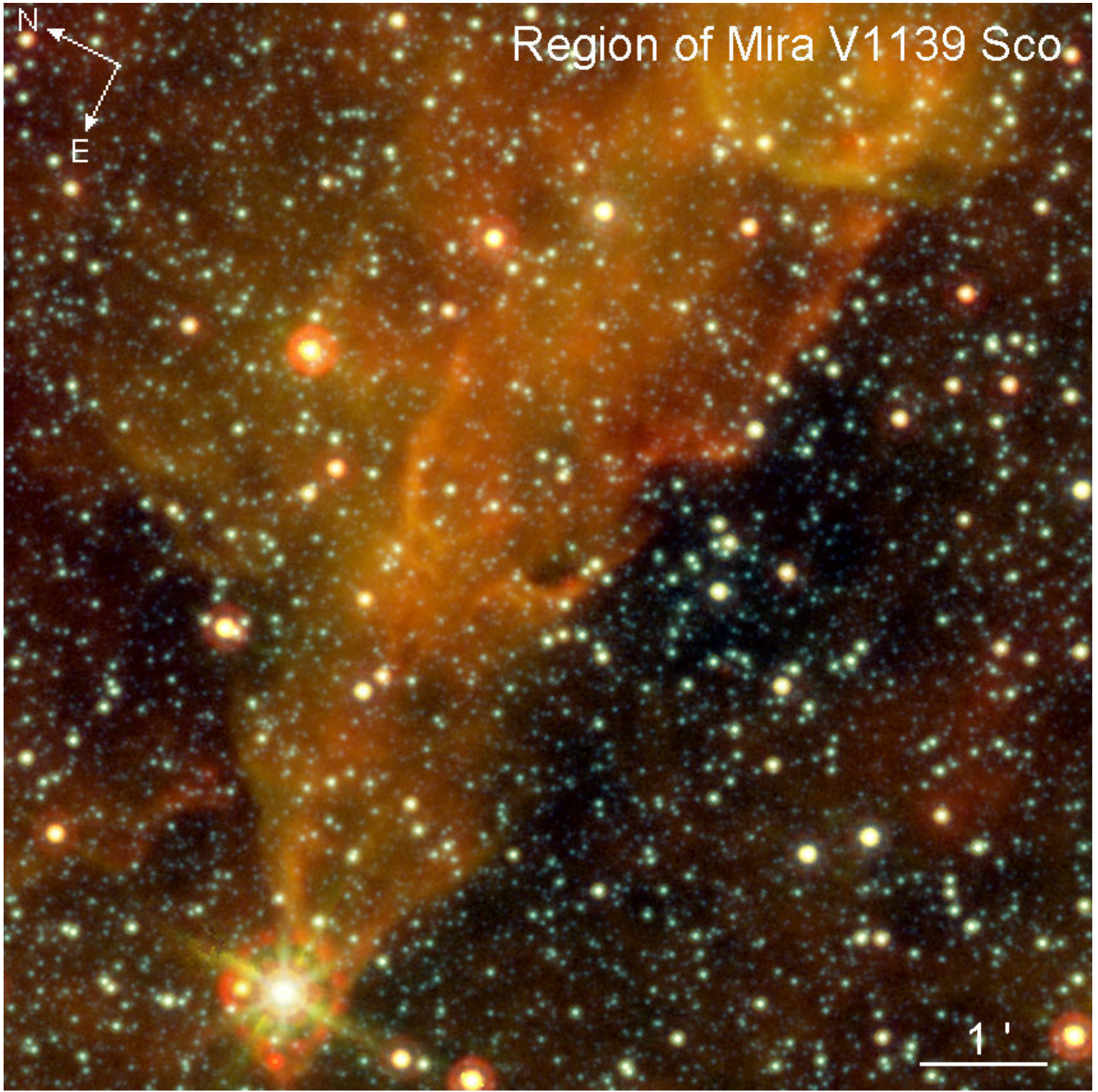

Figure 9